\documentclass[a4paper,11pt]{article}
\usepackage{pos}

\title{In medium evolution of the energy-energy correlators}

\author[a]{Weiyao Ke}

\affiliation[a]{Key Laboratory of Quark and Lepton Physics (MOE) \& Institute of Particle Physics, Central\\
China Normal University, Wuhan 430079, China}
\emailAdd{weiyaoke@ccnu.edu.cn}

\abstract{In this work, we present a first-principles analysis of the scale evolution of the two-point energy-energy correlator (EEC) for quark and gluon jets propagating through QCD matter. The EEC is a jet substructure observable that encodes the angular distribution of energy flow within jets and has proven valuable for precision tests of QCD in elementary collisions. Extending this framework to reactions with nuclei, we derive a factorized description of the in-medium EEC using Soft Collinear Effective Theory with Glauber gluon interactions, allowing systematic inclusion of medium-induced interactions. Working in the opacity expansion, we compute the medium-modified quark and gluon jet functions at one loop and perform leading-logarithmic resummation of large-scale logarithms. We find an experimentally accessible kinematic regime where medium effects manifest directly through medium-induced corrections to the anomalous dimensions, providing a transparent probe of in-medium dynamics. We test the theoretical framework by comparing our analytic predictions with EEC measurements in p–Pb collisions and discuss its implications for small collision systems (O-O) and the EIC.}

\FullConference{The 33rd International Workshop on Deep Inelastic Scattering and Related Subjects (DIS2026)\\
4 - 8 May 2026\\
Bologna, Italy\\}

\newcommand{\J}{\mathcal{J}}
\newcommand{\Hc}{\mathcal{H}}
\newcommand{\as}{\alpha_s}
\newcommand{\Leff}{L_{\rm eff}}
\newcommand{\rhoeff}{\rho_{\rm eff}}
\newcommand{\thLPM}{\theta_{\rm LPM}}
\newcommand{\mumed}{\mu_{\rm m}}
\hypersetup{pdftitle={In medium evolution of the energy-energy correlators},
 pdfauthor={Weiyao Ke},
 pdfsubject={DIS2026 contributed-talk proceedings}}
 
\begin{document}
\maketitle

\section{Introduction}

Energy correlators resolve jet substructure through energy-weighted correlations of jet constituents. At an angular separation $\theta$, the scale at which they resolve the underlying dynamics is $\mu_\theta=p_T\theta$, where $p_T\equiv p_T^{\rm jet}$. In elementary collisions, the small-angle EEC is governed by the time-like anomalous dimensions of QCD~\cite{Basham:1978,Dixon:2019}. In the limiting scenario of a high-energy jet passing through a finite-size medium, the medium modifies the time-like scale evolution of the parton and introduces scales associated with its length, screening, and transverse momentum broadening. The EEC can therefore test how jet scale evolution changes in QCD matter.

We summarize the calculation of Ref.~\cite{Ke:2026}, based on Soft Collinear Effective Theory with Glauber gluon interactions (SCET$_{\rm G}$)~\cite{Ovanesyan:2011,Ovanesyan:2012}. For energetic jets traversing a small or dilute medium, the first-order opacity correction modifies the EEC anomalous dimensions in a perturbative angular window. This extends the in-medium evolution analysis of Ref.~\cite{Ke:2024} to energy correlators, connecting their angular dependence to integrated medium properties.

\section{Factorization and the thin-medium regime}

The operator definition of the EEC involves two insertions of the energy-flow operator,
\begin{align}
 \Sigma(\hat n_1,\hat n_2)&=\langle\widehat{\mathcal E}(\hat n_1)\widehat{\mathcal E}(\hat n_2)\rangle,
 &\widehat{\mathcal E}(\hat n)&=\lim_{r\to\infty}r^2\int_0^\infty dt\,n_iT^{0i}(t,r\hat n).
 \label{eq:operator}
\end{align}
Here $\widehat{\mathcal E}$ is the energy-flow operator and $T^{0i}$ are the momentum density components of the energy-momentum tensor. $\hat{n}_1$ and $\hat{n}_2$ are unit vectors specifying the directions in which the energy flow is measured inside the jet.
For reconstructed jets, we use the per-jet distribution at nonzero angle,
\begin{align}
 C_{AB}(\theta;p_T,y)
 &=\left(\frac{d\sigma_{AB}^{\rm jet}}{dp_Tdy}\right)^{-1}
   \frac{d\Sigma_{AB}}{d\theta^2dp_Tdy}\,.
 \label{eq:observable}
\end{align}
We consider anti-$k_T$ jets with radius $R$ and the collinear region $\theta\ll R$. The energy-weighted cross section factorizes into parton distributions, a hard partonic cross section, and a semi-inclusive EEC jet function,
\begin{align}
 \frac{d\Sigma_{AB}}{d\theta^2dp_Tdy}
 &=\sum_{a,b,i,X}\int dx_a\,dx_b\,\frac{dz_J}{z_J}\,
 f_{a/A}(x_a,\mu)f_{b/B}(x_b,\mu)\nonumber\\
 &\quad\times\Hc_{ab\to iX}\!\left(\frac{p_T}{z_J},y,\mu\right)
 \J_{{\rm EEC},i}^{\rm vac+med}(\theta;z_J,p_T,R,\mu).
 \label{eq:factorization}
\end{align}
The momentum fraction $z_J$ relates the reconstructed jet to its parent parton. At one loop, out-of-cone radiation can be separated from the exclusive EEC jet function,
\begin{align}
 \J_{{\rm EEC},i}^{\rm vac+med}(\theta;z_J,p_T,R,\mu)
 =\sum_j\bigl[H_{ji}+\Delta H_{ji}\bigr]\,
          \J_{\rm EEC,j}^{\rm vac+med}(\theta;p_T,\mu),
 \label{eq:matching}
\end{align}
with products expanded to the matching accuracy. The hard-collinear coefficients $H_{ji}(z_J,p_T,R,\mu)$ describe the jet energy fraction and flavor and obey time-like DGLAP evolution~\cite{Kang:2016}. Their medium corrections include collisional energy loss, out-of-cone radiation, and flavor conversion. In this work, the medium properties are first averaged over production points and jet trajectories in each centrality class. The exclusive jet function $\J_{\rm EEC,j}(\theta;p_T,\mu)$ describes energy correlations inside the jet and has been expanded in $\theta/R$.

The theoretical calculation is organized as an expansion in the medium opacity $\chi_i$, i.e., the mean number of collisions between the jet parton $i$ and the medium. 
For screened Glauber exchange,
\begin{align}
 \chi_i&=\int dz\,\rho_G(z)\int\frac{d^2q_\perp}{(2\pi)^2}
 \frac{g_s^2C_i}{(q_\perp^2+m_D^2)^2}
 \sim\frac{\as C_i\rho_G L}{m_D^2}\,, \quad \rho_G(z)=\sum_b n_b(z)\frac{g_s^2C_b}{d_A}.
 \label{eq:opacity}
\end{align}
Here $C_q=C_F$, $C_g=C_A$, $d_A=N_c^2-1$, and $n_b$ is the number density of medium constituents in color representation $b$ in thermal equilibrium. In a thermal plasma, $m_D^2=(1+N_f/6)g_s^2T^2$. 

Corrections to the energy flow from medium-induced parton radiation affect the energy correlators. For a thin medium, collinear radiation is coherent over the entire medium and is characterized by $p^2\sim p_T/L$. Thus, the medium-induced radiative corrections to the EEC are organized as an expansion in powers of the small parameter 
\begin{align}
\alpha_s v = \alpha_s \chi_i \times \frac{m_D^2}{p_T/L} = \frac{\alpha_s^2 C_i \rho_G L}{p_T/L}\,.
\end{align}
The calculation assumes $\chi\sim 1$ and $v\ll 1$; thus, we require
\begin{align}
 \frac{p_T}{L}\gg \alpha_s \rho_G L \sim \ m_D^2,
 \label{eq:hierarchy}
\end{align}
and compute the radiation correction through $\mathcal O(\as\chi)$ or  $\mathcal O(\as v)$. This regime is particularly relevant to small collision systems and cold nuclear matter. 

\section{Medium corrections to the matching and evolution}

\subsection{Vacuum evolution and the fixed-order medium correction}

The exclusive jet function contains non-contact terms, where the energy weights act on different daughters of a splitting, and contact terms, where both act on the same daughter. The latter involve daughter EECs weighted by $x^2$ or $(1-x)^2$ and generate scale evolution. Evolving upward from $\mu_\theta$ to $\mu_R=2p_T\tan(R/2)$, the vacuum equation is
\begin{align}
 \frac{\partial\J_{\rm EEC,i}^{\rm vac}}{\partial\ln\mu^2}
 &=-\frac{\as(\mu)}{4\pi}\sum_j\gamma_{ji}^{\rm vac}(3)\J_{\rm EEC,j}^{\rm vac},
 &\gamma_{ji}^{\rm vac}(N)&=-2\int_0^1dx\,x^{N-1}P_{ji}^{(0)}(x).
 \label{eq:vac-rge}
\end{align}
The regularized splitting kernels $P_{ji}^{(0)}$ include virtual terms. The two-point EEC probes the moment $N=3$. In the quark--gluon basis, with matrix elements $\Gamma_{ij}=\gamma_{ji}$,
\begin{align}
 \boldsymbol\Gamma^{\rm vac}(3)
 &=\begin{pmatrix}
  \frac{25}{6}C_F & -\frac{7}{6}C_F\\[2pt]
  -\frac{14}{15}N_fT_F & \frac{14}{5}C_A+\frac{2}{3}N_f
 \end{pmatrix}\,,\quad \J_{\rm EEC,i}^{\rm vac}(\mu_\theta)
 =\frac{\as(\mu_\theta)}{2\pi\theta^2}A_i
   \left[1+\mathcal O\!\left(\frac{\Lambda_{\rm QCD}}{p_T\theta}\right)\right],
 \label{eq:vac-boundary}
\end{align}
where $A_q=3C_F/4$, $A_g=7C_A/10+N_fT_F/10$, and $T_F=1/2$. Evolving this boundary condition resums the leading collinear logarithms.

At first order in opacity, the non-contact medium correction follows from the SCET$_{\rm G}$ splitting spectra~\cite{Ovanesyan:2012},
\begin{align}
 J_{\rm EEC,i}^{\rm nc, med}(\theta)
 =\sum_{(jk)}\int dx\,d^2k_\perp\,d^2q_\perp\,
 \frac{dN_{i\to jk}^{\rm med}}{dx\,d^2k_\perp\,d^2q_\perp}
 x(1-x)\,\Theta_{jk}^{<R}
 \delta\!\left(\theta^2-\frac{k_\perp^2}{x^2(1-x)^2p_T^2}\right).
 \label{eq:noncontact}
\end{align}
The spectrum includes the integration over the scattering position, and $\Theta_{jk}^{<R}$ restricts both daughters to the jet. The Landau--Pomeranchuk--Migdal (LPM) interference phase compares the parton formation time to the medium length,
\begin{align}
 1-\cos\!\left[\frac{k_\perp^2L}{2x(1-x)p_T}\right]
 =1-\cos\!\left[4\pi x(1-x)\frac{\theta^2}{\thLPM^2}\right],
 \qquad \thLPM=\sqrt{\frac{8\pi}{p_TL}}.
 \label{eq:lpm}
\end{align}
Therefore, the characteristic angle at which the fixed-order correction changes is $\theta\sim\thLPM$. Its leading magnitude contains a Coulomb logarithm from the $1/q_\perp^4$ tail of screened scattering. For an exponentially decaying medium, the analytic result suggests a scaling
\begin{align}
 \J_{\rm EEC,i}^{\rm nc, med}(\theta)
 \sim\as^2\rhoeff\Leff^3
 \ln\!\left(\frac{p_T}{\Leff m_D^2}\right)
 F_i\!\left(\frac{\theta}{\thLPM}\right),
 \label{eq:coulomb}
\end{align}
with channel-dependent color factors included in $F_i$. This scaling is supported by numerical calculations for a realistic QGP medium~\cite{Ke:2026}.

\subsection{Medium-induced anomalous dimensions}

Emissions with formation times much longer than $L$ are coherent over the entire medium. Their contribution modifies the power-law exponent of the EEC in the window $\Lambda_{\rm QCD}/p_T \ll\theta\ll\thLPM$. 
To extract this contribution, we calculate the contact terms in dimensional regularization and expand in $m_D^2/(p_T/L)$. Renormalizing their poles gives
\begin{align}
 \boldsymbol\Gamma^{\rm med}(N)
 =&w_{\rm med}\begin{pmatrix}
 2(N-1)C_FC_A+C_F^2 & -C_F^2\\
 -2N_fT_FC_F & 2(N-1)C_A^2+2N_fT_FC_F
 \end{pmatrix}\nonumber\\
 w_{\rm med}=&4\pi\as(\mu)\,
 \frac{\rhoeff\Leff}{2p_T/\Leff}\ll1.
 \label{eq:medium-gamma}
\end{align}
The diagonal entries grow linearly with the moment $N$, while the off-diagonal entries describe quark--gluon mixing. 
The moment $N$ dependence of the diagonal entries also motivates ratios of higher-point correlators to the EEC as further tests of in-medium evolution.
The medium dependence is encoded by two path-integrated quantities $\rho_{\rm eff}$ and $L_{\rm eff}$, defined by
\begin{align}
 \int_0^\infty dz\,z\rho_G(z)\ln\frac{z}{\Leff}=0,
 \qquad
 \rhoeff=\frac{1}{\Leff^2}\int_0^\infty dz\,z\rho_G(z)\,.
 \label{eq:effective-medium}
\end{align}
These definitions apply to an arbitrary density profile along the trajectory. The dependence on integrated properties follows from the coherence of the radiation over the medium; the anomalous dimension does not resolve the local density profile at this order.

\begin{figure}[t]
 \centering
 \includegraphics[width=.9\textwidth]{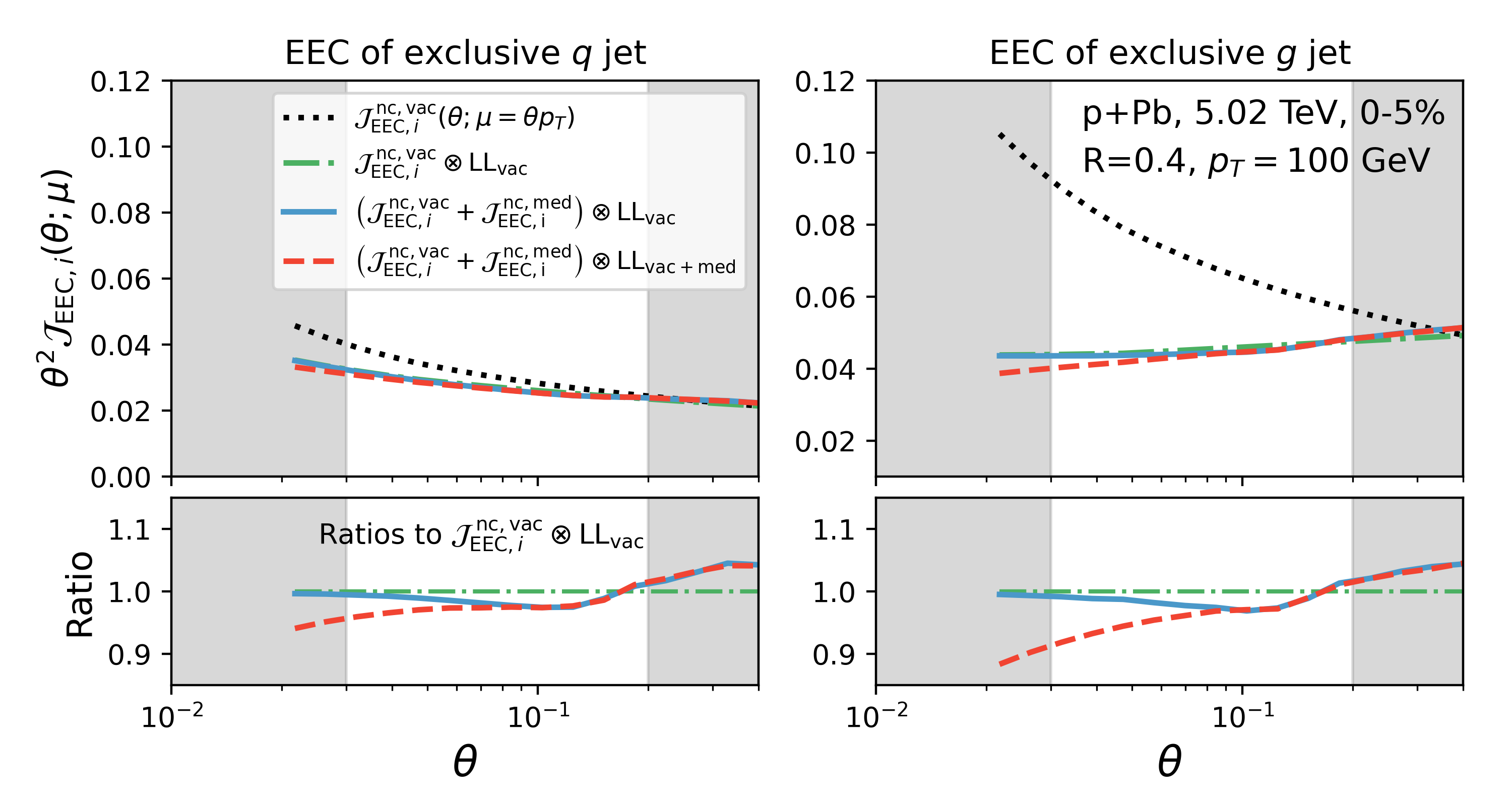}
 \caption{Exclusive quark- and gluon-jet EECs for $p_T=100$ GeV, $R=0.4$, and $0$--$5\%$ $p$--Pb collisions at $\sqrt{s_{NN}}=5.02$ TeV. Black: vacuum boundary conditions; green: vacuum LL evolution; blue: adding the non-contact medium correction; red: including medium-modified LL evolution. Lower panels show ratios to vacuum LL. Shaded regions indicate potentially large hadronization or large-angle corrections. From Ref.~\cite{Ke:2026}.}
 \label{fig:exclusive}
\end{figure}

The evolution switches from vacuum-like to medium-modified evolution around $\mumed\sim\sqrt{p_T/\Leff}$. For $\mu_\theta\ll\mumed\ll\mu_R$, we evolve with $\boldsymbol\Gamma^{\rm vac}+\boldsymbol\Gamma^{\rm med}$ below $\mumed$, include the finite medium matching corrections there, and continue with vacuum evolution up to $\mu_R$. For $\mu_\theta\gtrsim\mumed$, the fixed-order medium correction enters the boundary condition and the subsequent collinear evolution is vacuum-like. This separates logarithmically enhanced coherent radiation from radiation that resolves the medium length.

Figure~\ref{fig:exclusive} illustrates the distinction. The non-contact correction depletes the EEC below the LPM angle and enhances it at larger angles. Medium-modified evolution produces additional suppression at small angles, especially for gluon jets because of their larger color factors. The $w_{\rm med}\propto \alpha_s\rhoeff\Leff^2/p_T$ dependence of the medium-induced anomalous dimension also implies weaker corrections at higher jet energies. 

\section{Phenomenology in small systems and implications for the EIC}

We compare the per-jet nuclear modification factor $R_{AB}(\theta)=C_{AB}(\theta)/C_{pp}(\theta)$ with preliminary ALICE measurements in minimum-bias $p$--Pb collisions~\cite{ALICE:2025} (for results submitted to journals, see Ref.~\cite{ALICE:2026}). Medium corrections are averaged over hydrodynamic profiles from iEBE-VISHNU~\cite{Shen:2016}. Besides the exclusive EEC modification, the calculation includes energy loss in the hard-collinear coefficients, which changes the quark--gluon mixture in a fixed reconstructed-$p_T$ bin.

\begin{figure}[t]
 \centering
 \includegraphics[width=0.6\textwidth]{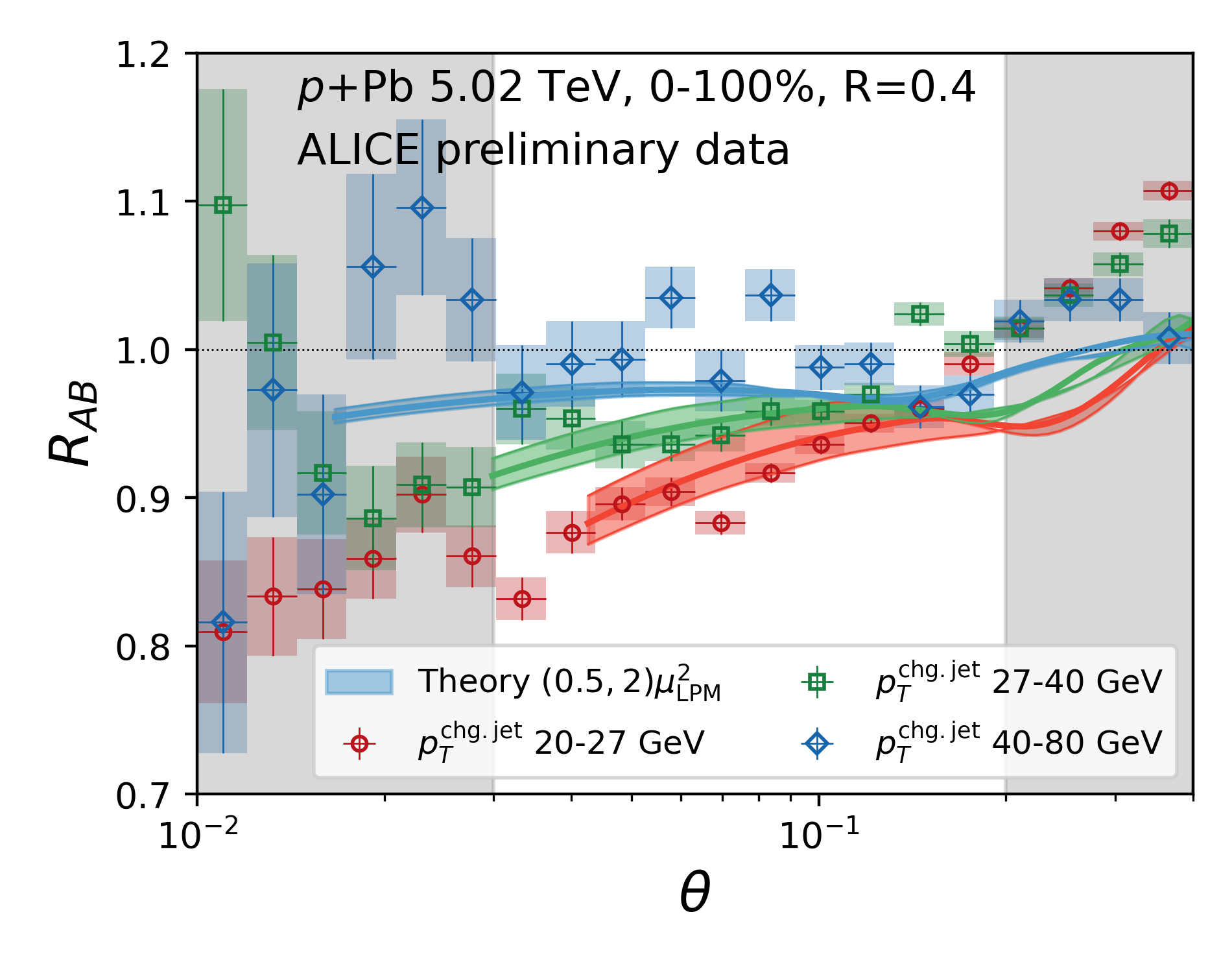}
 \caption{Per-jet EEC modification in $0$--$100\%$ $p$--Pb collisions at $\sqrt{s_{NN}}=5.02$ TeV and $R=0.4$, compared with preliminary ALICE data~\cite{ALICE:2025} (for results submitted to journals, see Ref.~\cite{ALICE:2026}). The red, green, and blue results correspond to charged-jet intervals $20$--$27$, $27$--$40$, and $40$--$80$ GeV. Theory bands vary the medium matching scale by a factor of two. The shaded angular regions are outside the main perturbative comparison window. From Ref.~\cite{Ke:2026}.}
 \label{fig:ppb}
\end{figure}

Figure~\ref{fig:ppb} shows suppression at small and intermediate angles and recovery at larger angles. The suppression weakens with increasing jet energy, consistent with the preliminary data. Within the window $\Lambda_{\rm QCD}/p_T \ll\theta\ll\thLPM$, this behavior is consistent with the medium-induced anomalous dimensions. Confirmation in a perturbatively controlled angular range would provide evidence for modified scale evolution in a small system. The experimental measurement uses charged jets. Here we approximate the charged-jet momentum by $p_T^{\rm ch}\simeq(2/3)p_T^{\rm full}$ and compare the full-jet EEC with the data. A complete charged-particle calculation requires track functions~\cite{Chen:2020}. At the lowest angles, hadronization becomes important, while at $\theta\sim R$, recoil, medium response, and corrections to the collinear approximation can affect the result. The comparison therefore emphasizes the intermediate perturbative window. The displayed matching-scale bands do not include all uncertainties in the medium parameters.

\begin{figure}[t]
 \centering
 \includegraphics[width=0.9\textwidth]{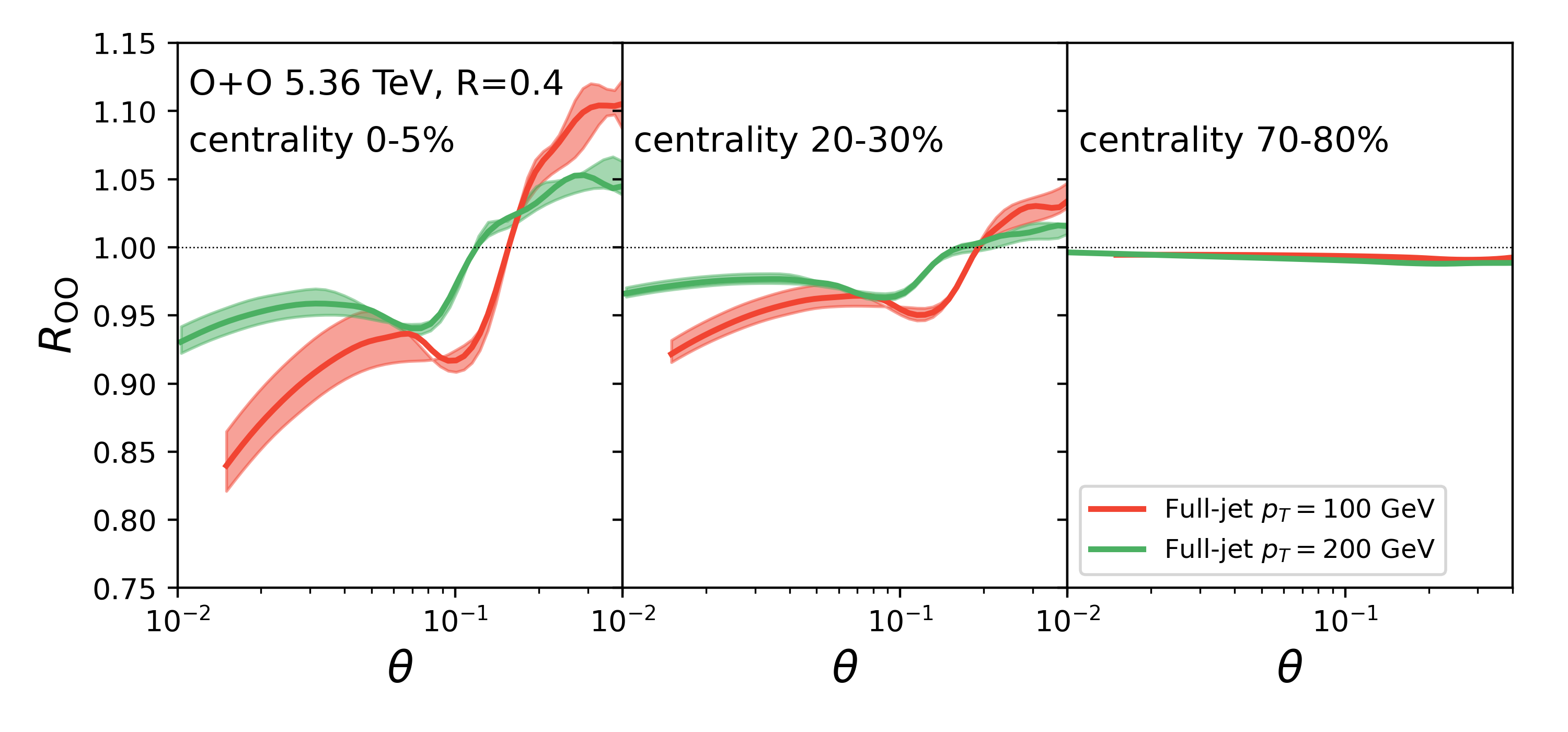}
 \caption{Full-jet EEC projections for O--O collisions at $\sqrt{s_{NN}}=5.36$ TeV and $R=0.4$. The panels show $0$--$5\%$, $20$--$30\%$, and $70$--$80\%$ centrality; red and green curves correspond to $p_T=100$ and $200$ GeV. The hydrodynamic and jet--medium parameters are the same as in the $p$--Pb calculation. From Ref.~\cite{Ke:2026}.}
 \label{fig:oo}
\end{figure}

Figure~\ref{fig:oo} presents O--O projections obtained with the same parameters, changing only the collision energy and nuclear species. The modification is stronger in central collisions and at lower jet energy, and decreases toward peripheral events. Measurements across $p$--Pb and O--O can test the dependence on $\rhoeff\Leff^2/p_T$ and the migration of the LPM angle with energy and medium size. 

\section{Summary}

At first order in opacity and for $p_T/L\gg\alpha_s \rho_G L \sim m_D^2$, radiation resolving the medium length modifies EEC matching around $\thLPM$, with a screening-regulated Coulomb logarithm. Coherent radiation modifies the quark and gluon anomalous dimensions for $\Lambda_{\rm QCD}/p_T\ll\theta\ll\thLPM$. Their dependence on integrated medium properties and jet energy provides a direct test of in-medium scale evolution. The preliminary $p$--Pb comparison and O--O projections motivate further measurements in small systems. The hierarchy $E/L\gg\hat \alpha_s \rho_G L$ is also relevant to energetic current jets in nuclear DIS. Energy correlators have been proposed as probes of cold nuclear matter~\cite{Devereaux:2025}; coherent scale evolution provides an additional mechanism to investigate at the EIC. Quantitative predictions require nuclear density profiles, DIS kinematics, and the screening scale of cold nuclear matter.

{\bf Acknowledgments:} W.K. is supported by the National Natural Science Foundation of China under Grant No.\ 12575140.

\end{document}